\documentclass[%
 reprint,
superscriptaddress,
 amsmath,amssymb,
 aps,
prl,
]{revtex4-2}

\usepackage{graphicx}
\usepackage{epstopdf}
\usepackage{dcolumn}
\usepackage{bm}
\usepackage{gensymb}
\usepackage{upgreek}
\usepackage{hyperref}
\usepackage{tabularx}
\usepackage{soul}
\usepackage[dvipsnames]{xcolor} 
\preprint{APS/123-QED}

\begin{document}
\setstcolor{blue}


\title{Elastoresistivity in the incommensurate charge density wave phase of BaNi$_2$(As$_{1-x}$P$_x$)$_2$}


\author{M. Frachet}
\thanks{mehdi.frachet@kit.edu}
\affiliation{Institute for Quantum Materials and Technologies, Karlsruhe Institute of Technology, 76021 Karlsruhe, Germany}

\author{P. W. Wiecki}
\affiliation{Institute for Quantum Materials and Technologies, Karlsruhe Institute of Technology, 76021 Karlsruhe, Germany}

\author{T. Lacmann}
\affiliation{Institute for Quantum Materials and Technologies, Karlsruhe Institute of Technology, 76021 Karlsruhe, Germany}

\author{S. M. Souliou}
\affiliation{Institute for Quantum Materials and Technologies, Karlsruhe Institute of Technology, 76021 Karlsruhe, Germany}

\author{K. Willa}
\affiliation{Institute for Quantum Materials and Technologies, Karlsruhe Institute of Technology, 76021 Karlsruhe, Germany}

\author{C. Meingast}
\affiliation{Institute for Quantum Materials and Technologies, Karlsruhe Institute of Technology, 76021 Karlsruhe, Germany}

\author{M. Merz}
\affiliation{Institute for Quantum Materials and Technologies, Karlsruhe Institute of Technology, 76021 Karlsruhe, Germany}
\affiliation{Karlsruhe Nano Micro Facility (KNMFi), Karlsruhe Institute of Technology, 76021 Karlsruhe, Germany}

\author{A.-A. Haghighirad}
\affiliation{Institute for Quantum Materials and Technologies, Karlsruhe Institute of Technology, 76021 Karlsruhe, Germany}

\author{M. Le Tacon}
\affiliation{Institute for Quantum Materials and Technologies, Karlsruhe Institute of Technology, 76021 Karlsruhe, Germany}

\author{A. E. B\"ohmer}
\affiliation{Institute for Quantum Materials and Technologies, Karlsruhe Institute of Technology, 76021 Karlsruhe, Germany}
\affiliation{Lehrstuhl für Experimentalphysik IV,  Fakultät für Physik und Astronomie, Ruhr-Universität Bochum, 44801 Bochum, Germany}

\date{\today}


\begin{abstract}

Electronic nematicity, the breaking of the crystal lattice rotational symmetry by the electronic fluid, is a fascinating quantum state of matter. Recently, BaNi$_2$As$_2$ has emerged as a promising candidate for a novel type of nematicity triggered by charge fluctuations. In this work, we scrutinize the electronic nematicity of BaNi$_2$(As$_{1-x}$P$_x$)$_2$ with $0 \leq x \leq 0.10$ using electronic transport measurements under strain. We report a large $B_{1g}$ elastoresistance coefficient that is maximized at a temperature slightly higher than the first-order triclinic transition, and that corresponds to the recently discovered tetragonal-to-orthorhombic transition \citep{Merz_PRB_2021}. The reported elastoresistance does not follow the typical Curie-Weiss form observed in iron-based superconductors but has a much sharper temperature dependence with a finite elastoresistance onsetting only together with a strong enhancement of the incommensurate charge density wave of the material. Consequently, the $B_{1g}$ elastoresistance and the associated orthorhombic distortion appears here as a property of this incommensurate charge density wave. Finally, we report and track the hysteretic behavior seen in the resistance versus strain sweeps and interpret its origin as the pinning of orthorhombic domains. Our results revise the understanding of the interplay between nematicity, charge density waves and structural distortions in this material.
\end{abstract}


\maketitle

\section{Introduction}

With the discovery of iron-based superconductors \citep{Hosono_2006}, electronic nematicity has emerged as a potential key ingredient for high-temperature superconductivity. Indeed, the observation of strong electronic nematic fluctuations at the optimal conditions for superconductivity suggests that such fluctuations might promote higher $T_\textrm{c}$ \citep{Chu_Science_2012, Kuo_Science_2016, Anna_PRL_2014, Fernandes_NaturePhysics_2014, Lederer_PRL_2015}. This view is supported by reports for nematicity in other unconventional superconductors as heavy fermions \citep{Ronning_Nature_2017, Okazaki_Science_2011} or cuprates \citep{Murayama_NatComm_2019, Daou_Nature_2010, Sato_NatPhys_2017, Choiniere_PRB_2015, Auvray_NatComm_2019, Ishida_JPSP_2020, Kivelson_Nature_1998}.

Nonetheless, the best-understood case is by far the one of the iron pnictides, where the electronic nematic fluctuations induce a tetragonal-to-orthorhombic phase transition at $T_{\textrm{s}}$. The nematic transition, when not coincident to, is closely followed by antiferromagnetic order at $T_{\textrm{N}}$ \citep{Nandi_PRL_2010, Ni_PRB_2008, Anna_PRB_2012}. From this empirical observation and theoretical considerations \citep{Fernandes_NaturePhysics_2014} anisotropic magnetic fluctuations are a leading candidate for the mechanism of nematicity in iron-based superconductors. However, beyond the iron pnictides case, and in particular in the absence of long-range magnetic order \citep{Meingast_PRB_2018,Massat_PNAS_2016, Chibani_npjQMat_2021, Anna_PRB_2013, Ghini_PRB_2021, Hosoi_PNAS_2016}, much remains to be understood about the mechanisms of electronic nematicity and its significance for superconductivity.

In this regard, BaNi$_2$As$_2$, that shares the same high temperature tetragonal structure as the intensively studied BaFe$_2$As$_2$, has recently attracted attention. In contrast to its iron-analogue, BaNi$_2$As$_2$ is superconducting below $T_\textrm{c} \approx 0.7~$K at ambient pressure \citep{Sefat_PRB_2009} and hosts two types of charge density waves (CDWs). Upon cooling, first appears an incommensurate charge density wave (I-CDW) that develops strongly at $T_{\textrm{I-CDW}} \approx 155~$K. A weaker diffuse signal can be tracked all the way up to room temperature \citep{Merz_PRB_2021, Raman_KIT_Preprint, Demsar_CommunicationPhysics_2022}. The exact nature of this I-CDW is currently under intense investigation. The reciprocal space pattern of the I-CDW superlattice peaks reported by x-ray diffraction experiments indicated that the charge modulation is unidirectional \citep{Lee_PRL_2019, Lee_PRL_2021, Merz_PRB_2021, Raman_KIT_Preprint}. Thus, it has been initially linked to a breaking of the crystal lattice rotational symmetry \citep{Lee_PRL_2019, Eckberg_2020}. Although, from a wide reciprocal lattice mapping it has been later suggested that, overall, the I-CDW is a biaxial, rotationally invariant state \citep{Lee_PRL_2021}, a clear rotational symmetry breaking has been recently observed within the I-CDW phase by high-resolution thermal expansion measurements \citep{Merz_PRB_2021, Meingast_InPreparation}. At a slightly lower temperature a commensurate uniaxial charge density wave (C-CDW) develops at the expense of the former \citep{Lee_PRL_2019, Lee_PRL_2021, Merz_PRB_2021, Raman_KIT_Preprint, Demsar_CommunicationPhysics_2022} and is associated to a first-order triclinic structural transition at $T_\textrm{tri} \approx 137~$K. This latter coincident transition is suppressed through numerous chemical substitutions at a critical value, $x_c$ \citep{Lee_PRL_2019,Lee_PRL_2021, Raman_KIT_Preprint, Kudo_arXiv_2017, Kudo_PRL_2012, Merz_PRB_2021, Sefat_PRB_2009}.

\begin{figure*}
    \centering
    \includegraphics[width=17.2cm]{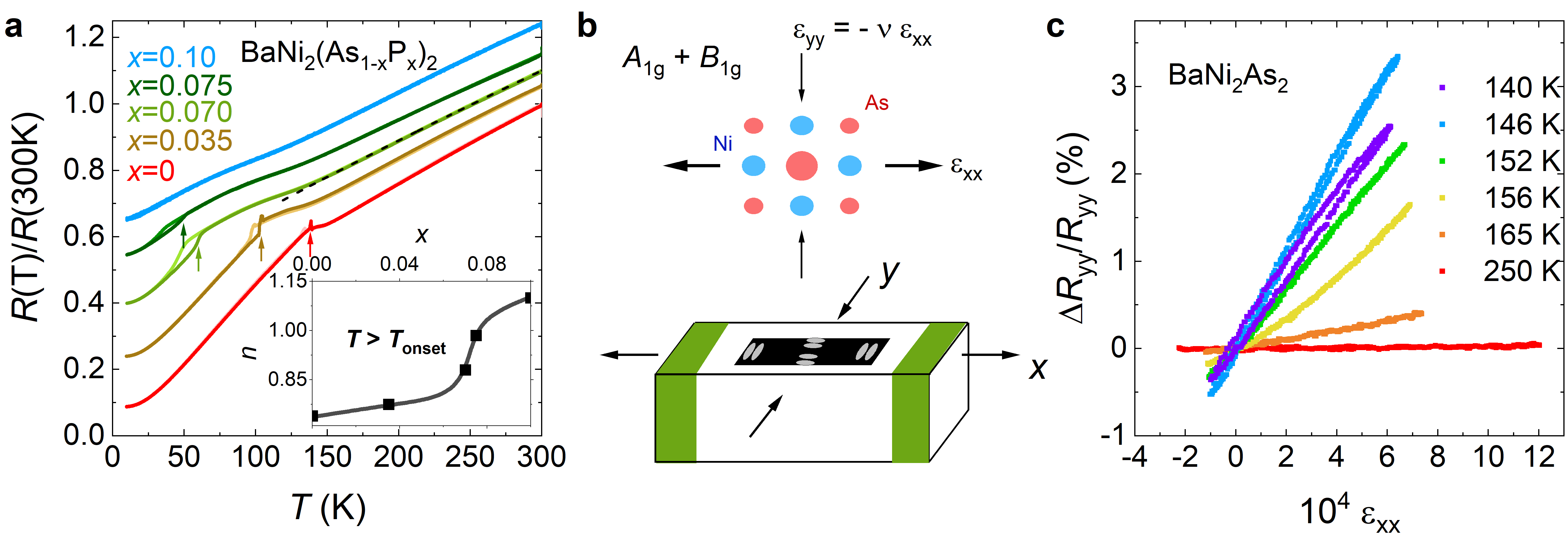}
    \caption{\textbf{Characterization and elastoresistance measurements of BaNi$_2$(As$_{1-x}$P$_x$)$_2$ single crystals.} \textbf{a} Normalized resistance, $R(T)/R(300\textrm{K})$, at the indicated $x$ values. Cooling and warming measurements are depicted as light and dark colors, respectively. The curves are vertically shifted for clarity. The arrows indicate the triclinic transition temperatures upon warming, and the dashed black line is a high temperature fit to $R=R_0 + A T^n $ for $T>T_{\textrm{onset}}$ ($T_{\textrm{onset}}$ being the onset temperature of the elastoresistance, see main text). The inset shows the evolution of the $n$ exponent on substitution level. \textbf{b} Schematic of an elastoresistance experiment: the sample (black rectangle) is glued on top of a piezo with its $[100]_{\rm{tet}}$ axis along the piezo poling direction, $x$. Application of a positive voltage to the piezo leads to a tensile strain along the poling direction and a compression in the orthogonal axis through Poisson effect. Both $R_{xx}$ (so-called longitudinal) and $R_{yy}$ (transverse) electrical resistances are measured, using the piezo frame notation. The corresponding $A_{1g}+B_{1g}$ strain state of the basal plane is also shown. \textbf{c} Representative resistance variation with strain in transverse geometry for $x=0$ at selected temperatures. The elastoresistance signal is strongest at $T^* \approx 146~$K while an hysteretic behavior is more pronounced at lower temperature.}
    \label{Fig1}
\end{figure*}

Despite the absence of static magnetism down to lowest investigated temperature \citep{Kothapalli_JP_CS_2010} possible indications for electronic nematicity have been reported. First, in Ba$_{1-x}$Sr$_{x}$Ni$_2$As$_2$ strain-dependent electrical transport measurements, namely elastoresistance, have been interpreted as a signature of a large $B_{1g}$ electronic nematic susceptibility upon approaching the triclinic phase transition in substituted samples \citep{Eckberg_2020}. Second, a continuous orthorhombic transition was recently found to precede the triclinic one in Ba(Ni$_{1-x}$Co$_{x}$)$_2$As$_2$ and BaNi$_2$(As$_{1-x}$P$_x$)$_2$ \citep{Merz_PRB_2021, Meingast_InPreparation}. Unlike in BaFe$_2$As$_2$ the in-plane orthorhombic axes are aligned with the tetragonal ones and the associated lattice distortion is much smaller. By analogy with the iron pnictides, it was proposed that this intermediate phase with broken rotational symmetry is a possible manifestation of charge-induced nematicity \citep{Merz_PRB_2021}.

 In addition, the superconducting $T_\textrm{c}$ has been found to sharply increase to $\approx 3~$K for substitution levels just above the suppression of the triclinic and C-CDW transition, but the origin of this enhancement is still debated. In Ba$_{1-x}$Sr$_{x}$Ni$_2$As$_2$ it has been associated with electronic nematic fluctuations \citep{Eckberg_2020}, while in BaNi$_2$(As$_{1-x}$P$_x$)$_2$ it has been attributed to an enhanced electron-phonon coupling through a lattice softening \citep{Kudo_PRL_2012}.

Thus, it is critically needed to assess the possible advent of electronic nematicity in these materials, and establish its interplay with the aforementioned lattice and electronic instabilities. In this work, we investigate the electronic nematicity of BaNi$_2$(As$_{1-x}$P$_x$)$_2$ with $0\leq x \leq 0.10$, a system for which the tetragonal-to-orthorhombic transition is established \citep{Merz_PRB_2021, Meingast_InPreparation}, using elastoresistance measurements. We report a large maximum of the $B_{1g}$-symmetric elastoresistance coefficient, $m_{12}-m_{11}$, that occurs, up to $x=0.075$, at the temperature of the orthorhombic transition, where the rotational symmetry is broken. Importantly, the elastoresistance onset corresponds to a strong increase in the I-CDW superlattice peak intensity that cannot be described by a Curie-Weiss-like temperature dependence. Thus, the anisotropic strain dependent electrical transport is a property of the I-CDW phase. Finally, a careful investigation of the hysteretic behavior of the resistance versus strain sweeps strongly suggests that the hysteresis originates from the pinning of orthorhombic domains.


\section{Results}

\subsection{Experimental details} 
We start by investigating the freestanding resistance of BaNi$_2$(As$_{1-x}$P$_x$)$_2$ single crystals in Fig.\ref{Fig1}a. A metallic behavior with residual-resistivity ratio (RRR) values in line with the literature \citep{Kudo_PRL_2012, Sefat_PRB_2009} is observed. For BaNi$_2$As$_2$ RRR $\approx 12$ \textit{i.e.} approximately the value found in BaFe$_2$As$_2$ \citep{Meingast_NatComm_2015}. In agreement with previous reports a sharp increase of electrical resistance occurs at $T_\textrm{tri} = 137~$K upon cooling in BaNi$_{2}$As$_{2}$, signaling the triclinic structural transition \citep{Kudo_PRL_2012, Sefat_PRB_2009}. The hysteresis indicates the first-order nature of the transition. Upon increasing P-substitution this transition is shifted towards lower temperature, the resistance upturn becomes a downturn, and the width of the thermal hysteresis increases. No such transition is observed for $x=0.10$, where cooling and warming measurements overlap, indicating a critical doping for the triclinic phase $x_{c} \approx 0.08$, in agreement with the literature \citep{Raman_KIT_Preprint, Kudo_PRL_2012, Meingast_InPreparation}.

At low temperature, and except at the highest ($x=0.10 > x_c$) P-content, the electrical resistance does not follow a $T^2$ temperature dependence \citep{Sefat_PRB_2009, Meingast_InPreparation}. At high enough temperatures the resistance is well described as $R=R_{0}+A \times T^{n}$ (see dashed blacked line in Fig.\ref{Fig1}a for $x=0.070$), with $A$ and $n$ being $x$-dependent. As seen in the inset of Fig.\ref{Fig1}a the exponent $n$ shows a significant increase across the triclinic critical point, and in particular $n \approx 1$, i.e. a linear-in-temperature resistance is observed within a narrow substitution range around $x_c$, whose origin is still unknown (more details on the fitting are given in SM section I).

In order to study the elastoresistance of BaNi$_2$(As$_{1-x}$P$_x$)$_2$, we induce a symmetry-breaking strain to our single crystals by gluing them on top of a piezo stack as visualized in Fig.\ref{Fig1}b, a technique initially used in strongly correlated systems in Ref. \citep{Chu_Science_2012}. This method allows to extract the elastoresistance coefficients defined as $m_{ii,jj}=1/R_{ii}(dR_{ii}/d\epsilon_{jj})$, where \textit{j} denotes the direction of the strain and $i=x$ (resp. $i=y$)  corresponds to longitudinal (resp. transverse) measurements with respect to the piezo poling axis, $x$. In the following, we use the Voigt notation for the elastoresistance coefficient, in particular $xx=1$ and $yy=2$, and formulate in terms of the irreducible representations of the high-temperature $D_{4h}$ tetragonal point group.

With the [100]$_{\rm{tet}}$ axis aligned to the piezo stack poling direction, as done in the following, the in-plane resistance anisotropy that develops under strain is proportional to the symmetry-resolved $B_\textrm{1g}$ elastoresistance coefficient, $m_{12}-m_{11}$,

\begin{equation}
  \left( \frac{\Delta R}{R}\right)_{xx}-\left( \frac{\Delta R}{R}\right)_{yy} = \left(m_{11}-m_{12}\right)\left(\epsilon_{xx}-\epsilon_{yy}\right)
  \label{Eq1}
\end{equation}

Since for sufficiently small anisotropy any potential electronic nematic order parameter is proportional to the resistance anisotropy, the associated electronic nematic susceptibility in the $B_{1g}$ channel is probed by the $m_{12}-m_{11}$ elastoresistance coefficient \citep{Chu_Science_2012, Chu_Science_2010, Eckberg_2020, Kuo_PRB_2013}.

\begin{figure}
    \centering
    \includegraphics[width=8.6cm]{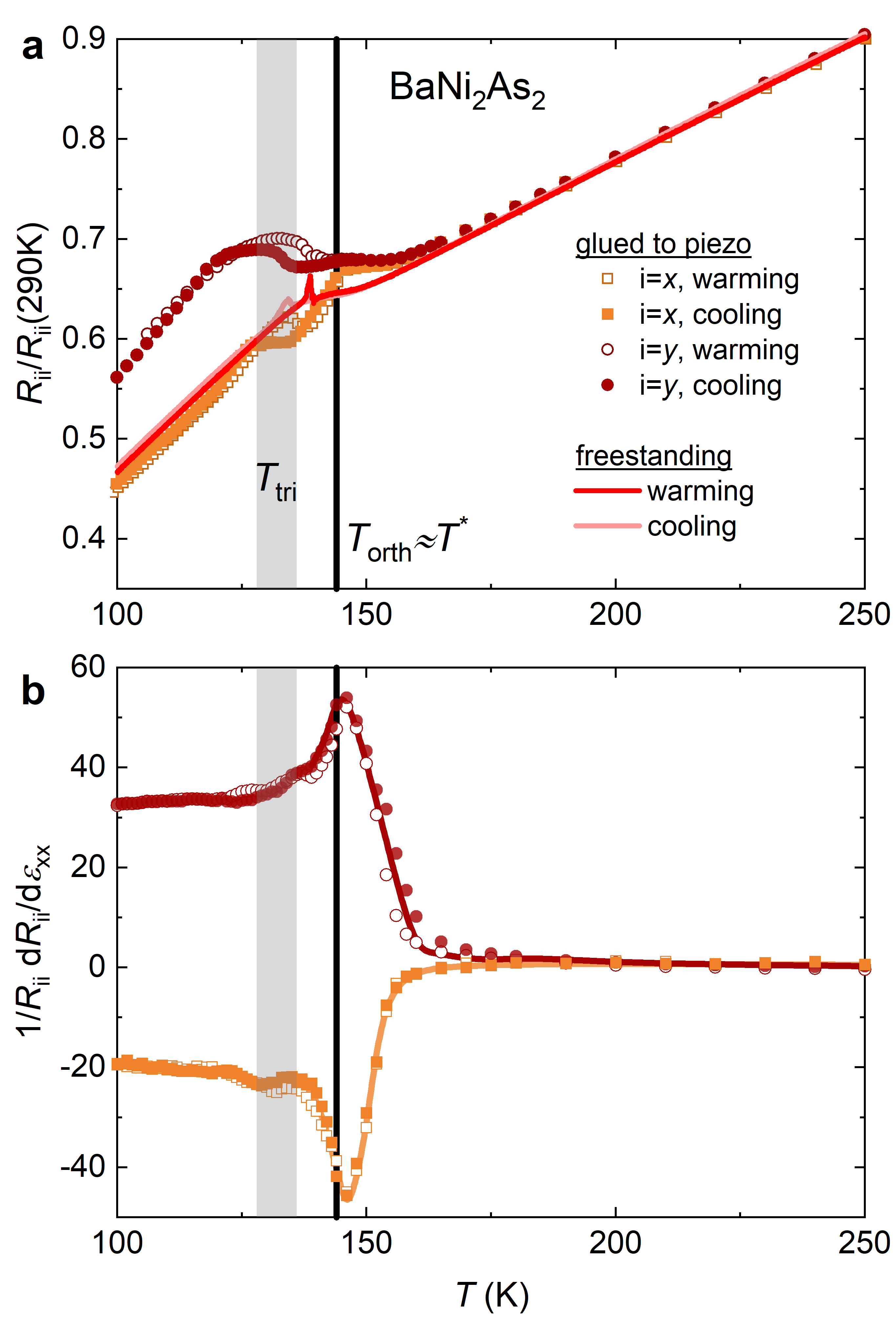}
    \caption{\textbf{Elastoresistance measurement of BaNi$_2$As$_2$.} \textbf{a} Normalized electrical resistances recorded with the sample [100]$_{\rm{tet}}$ axis glued along the poling direction of the piezo. Longitudinal ($i=x$, orange squares) and transverse measurements ($i=y$, dark red circles) correspond to resistances measured along and perpendicular to the piezo poling axis, respectively. No voltage is applied to the piezo. For comparison the freestanding resistance from Fig.\ref{Fig1}a is added (dark and light red lines for warming and cooling measurements, respectively). \textbf{b} Corresponding slopes of the normalized resistance versus strain variation, $1/R_{ii}  (dR_{ii}/d{\epsilon_{xx})}$, obtained through sweeping the piezo voltage at fixed temperatures. The filled (respectively empty) symbols correspond to cooling (resp. warming) measurements. The gray shaded area shows the temperature range of the triclinic structural transition upon cooling, while the vertical line denotes $T_\textrm{orth}$, \textit{i.e}. the orthorhombic transition temperature as determined by thermal expansion \citep{Merz_PRB_2021, Meingast_InPreparation}. The temperature of the maximum of the elastoresistance, $T^*$, is consistent with $T_{\rm{orth}}$. Lines are guide to the eye.}
    \label{Fig2}
\end{figure}

\subsection{Elastoresistance of BaNi$_2$As$_2$}
A typical example of raw data in BaNi$_2$As$_2$ in the transverse geometry is shown in Fig.\ref{Fig1}c. The corresponding detailed temperature dependence of the elastoresistance measurement is reported in Fig.\ref{Fig2}. 

First, we show the normalized resistances as a function of temperature with the sample being glued to the piezo, see Fig.\ref{Fig2}a. While both longitudinal and transverse directions (squares and circles, respectively) overlap at high enough temperature, a clear discrepancy appears below $T~\approx 145~$K, that corresponds to the second-order orthorhombic transition temperature, $T_\textrm{orth}$, as determined by thermal expansion measurements \citep{Merz_PRB_2021, Meingast_InPreparation}. This observation evidences that, within the orthorhombic state, the sample is, at least partially, detwinned through the anisotropic thermal expansion of the piezo (see SM section II) and the resulting strain.

\begin{figure*}
    \centering
    \includegraphics[width=17.2cm]{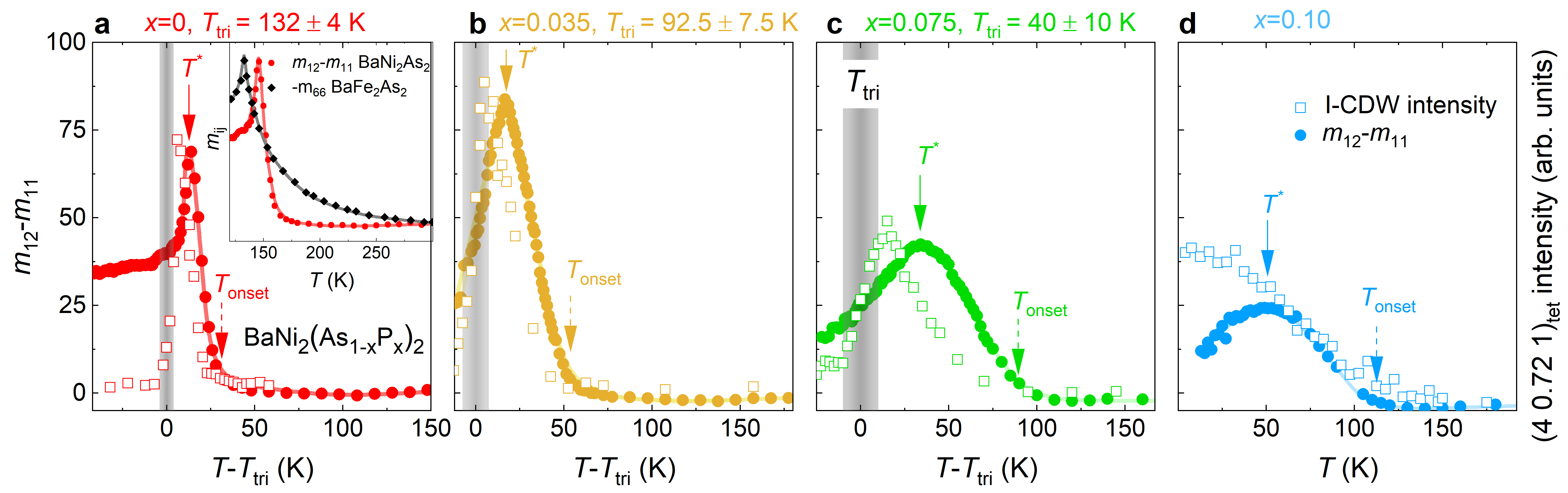}
    \caption{\textbf{Evolution of $m_{12}-m_{11}$ with P-concentration in BaNi$_2$(As$_{1-x}$P$_x$)$_2$.} \textbf{a-d} $B_{\textrm{1g}}$ symmetry-resolved $m_{12}-m_{11}$ elastoresistance coefficient (filled circles, left axis) together with the integrated intensity of the I-CDW satellite at $Q_{\rm{I-CDW}}=(4~0.72~1)_{\rm{tet}}$ (empty squares, right axis, reproduced from Ref.\citep{Raman_KIT_Preprint}), for different $x$ values as indicated. Both quantitities are shown upon cooling, except for the $x=0.10$ x-ray diffraction (XRD) data, and plotted as a function of $T-T_{\textrm{tri}}$, where the triclinic transition temperature is measured upon cooling. No triclinic phase is observed at $x=0.10$ and thus the data are as a function of $T$ in \textbf{d}. The gray area represents the width of the triclinic transition as determined from the elastoresistance samples (see SM section V for details). Note that elastoresistance and XRD measurements are performed on different samples that share similar triclinic transition temperatures, except for $x=0.075$ where the XRD sample has a slightly lower P-content and accordingly higher $T_{\textrm{tri}} \approx 55~$K. The arrows indicate the onset temperature of $m_{12}-m_{11}$, $T_\textrm{onset}$, and the temperature of the elastoresistance maximum, $T^*$. In the inset of \textbf{a} the $B_{2g}$-symmetric $-m_{66}$ elastoresistance coefficient of BaFe$_2$As$_2$ from Ref. \citep{Palmstrom_PRB_2017}  (black diamonds, vertically scaled) is compared to the $m_{12}-m_{11}$ coefficient of BaNi$_2$As$_2$. Lines are guide to the eye.}
    \label{Fig3}
\end{figure*}

The normalized freestanding resistance (dark and light red lines for the warming and cooling measurement, respectively), reproduced from Fig. \ref{Fig1}a, also overlaps with that of the sample glued to the piezo down to $\approx 160~$K. However, at lower temperatures a difference appears which can be attributed to the differential thermal expansion of the piezo and BaNi$_2$As$_2$ together with the emergence of a finite elastoresistance. The similar resistivity values observed in the longitudinal and transverse channels down to $T_\textrm{orth}$ are also in line with the mainly in-plane isotropic strain arising from the thermal expansion mismatch (see details in SM section II) and points to a finite in-plane symmetric elastoresistance.

While the gluing induces a broadening of the triclinic transition as seen by electrical transport, a thermal hysteresis is still clearly observed in these strained conditions and the transition temperature is not substantially shifted. Finally, note that the sign of the resistance anisotropy in the orthorhombic state is similar to the one in BaFe$_2$As$_2$, \textit{i.e.} the smaller in-plane orthorhombic axis (aligned with the $y$ piezo axis) is the one with the higher resistivity \citep{Chu_Science_2010, Meingast_NatComm_2015}.

The linear slopes of the resistance versus strain sweeps, $1/R_{ii}  (dR_{ii}/d{\epsilon_{xx}})$, are reported in Fig.\ref{Fig2}b. Note that they are extracted during the same temperature cycle as the corresponding resistances of Fig.\ref{Fig2}a, upon applying a voltage to the piezo stack at fixed temperatures. At high temperature no response to strain is seen in any channel, in agreement with the overlap of the respective electrical resistances. However, below $T_\textrm{onset} \approx 160~$K an elastoresistance signal develops sharply in both directions and peaks at $T^* = 145 \pm 2 ~\textrm{K}\approx T_\textrm{orth} $, with an opposite sign along the two directions. Thus, the $B_{1g}$ symmetric $m_{12}-m_{11}$  elastoresistance coefficient extracted from the difference of the longitudinal and transverse measurements is maximum at $T^* \approx T_\textrm{orth}$ (see Eq.\ref{Eq1} and Fig.\ref{Fig3}a; in the followings $T^*$ is formally defined as the temperature of the $m_{12}-m_{11}$ maximum). Notably, no strong feature appears at the triclinic transition temperature. This bring us to our first important result: the $B_{1g}$-symmetric $m_{12}-m_{11}$ maximum occurs at the orthorhombic transition rather than at the triclinic one, in contrast to what has been previously reported for Sr-substituted samples \citep{Eckberg_2020}. This is fully consistent with the 4-fold symmetry breaking that occurs at the orthorhombic transition \citep{Merz_PRB_2021, Meingast_InPreparation} and the absence of thermal hysteresis in the elastoresistance response, in particular around $T^*$ (see also additional measurements in SM section III).  \\


\subsection{Evolution with P-concentration}
The $m_{12}-m_{11}$ elastoresistance coefficient at the different P-contents investigated is shown in Fig.\ref{Fig3}b-d, as a function of $T-T_{\textrm{tri}}$, where the triclinic transition temperature and the elastoresistance coefficient are extracted upon cooling. A maximum of the elastoresistance is found up to the highest concentration investigated at temperature exceeding the triclinic transition. Rather, up to $x=0.075$, $T^*$ is in good agreement with $T_\textrm{orth}$ as determined by thermal expansion. With increasing substitution level, $T^*$ decreases smoothly towards $T^*\approx 50~$K for $x=0.10$ where the associated elastoresistance maximum becomes weaker and significantly broader. In particular, there is no enhancement of $m_{12}-m_{11}$ associated to the enhanced superconducting $T_\textrm{c}$ in the absence of the triclinic structure for $x=0.10$. Note that for $x=0.10$ no orthorhombic distortion is observed by high-resolution thermal expansion, which shows however a clear signature of a (different) first-order transition \citep{Meingast_InPreparation}. This is most likely not directly related to the elastoresistance maximum since, as for lower P-contents, no evidence for a thermal hysteresis is observed. We discuss this particular case later.

A fundamental aspect is the comparison between the temperature dependences of the $m_{12}-m_{11}$ elastoresistance coefficient and of the I-CDW superlattice peak intensity recorded at $Q_{\rm{I-CDW}}=(4~0.72~1)_{\rm{tet}}$ (empty squares, reproduced from Ref. \citep{Raman_KIT_Preprint}). Using samples from the same batches across the entire substitution range investigated, we find that both quantities onset at very similar temperatures. The elastoresistance coefficient is negligible above $T_\textrm{I-CDW}$, that corresponds to a strong increase in the I-CDW superlattice peak intensity. Thus, a large $m_{12}-m_{11}$ coefficient is a property of the incommensurate charge density wave phase and the associated fluctuations.

Finally, the maximum value of the $m_{12}-m_{11}$ elastoresistance coefficient displays a non-monotonic dependence on P-concentration and is highest at $x=0.035$, an observation that is reminiscent of Ba$_{1-x}$Sr$_{x}$Ni$_2$As$_2$ \citep{Eckberg_2020}. While this might be an artefact of strain transmission, one possible alternative scenario, that remains to be investigated, is a stronger orthorhombic distortion at this particular substitution content.

\begin{figure}
    \centering
    \includegraphics[width=8.6cm]{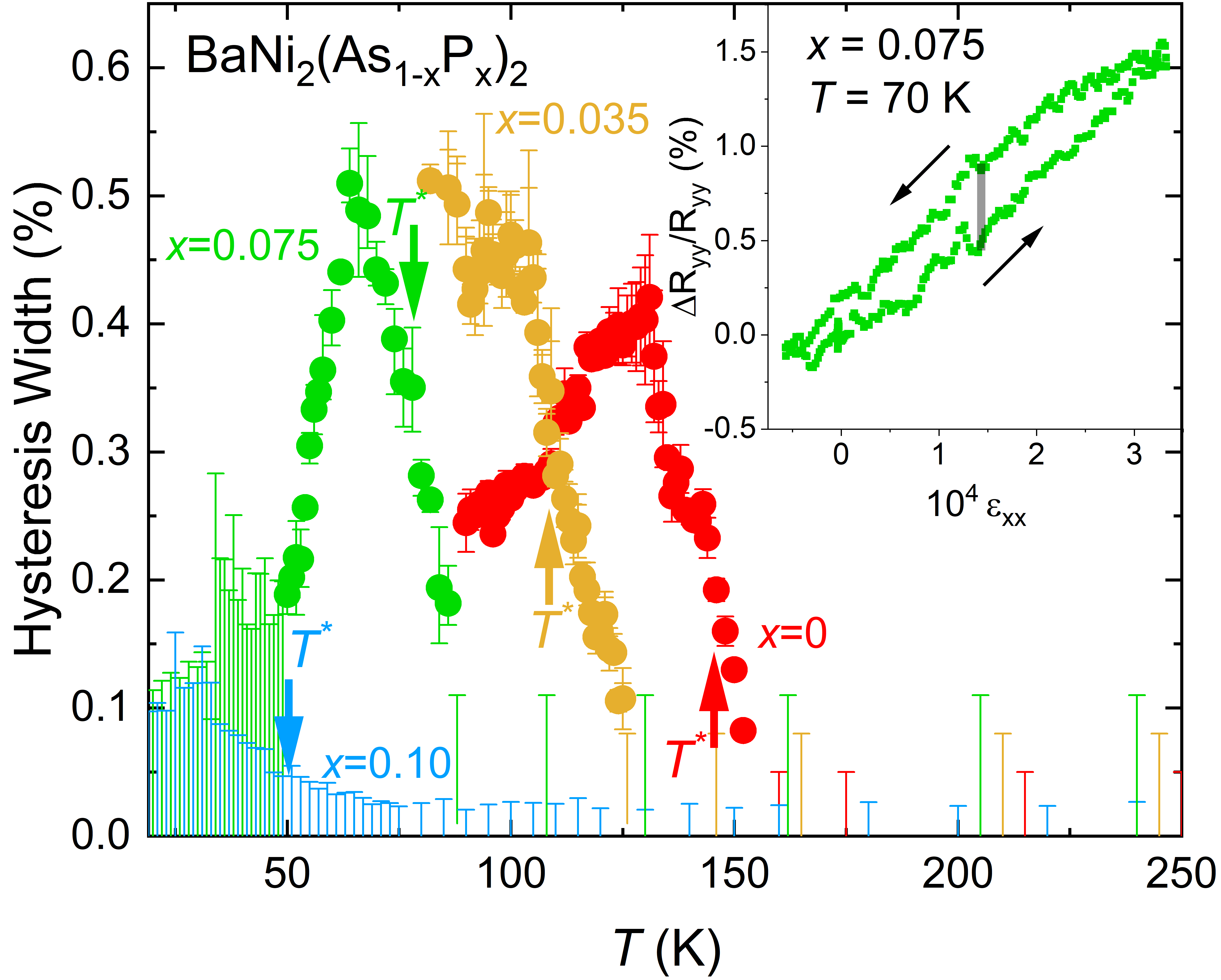}
    \caption{\textbf{Elastoresistance hysteresis as a function of P-substitution.} Temperature dependence of the hysteresis width for the indicated compositions. The inset shows the definition of the hysteresis width, \textit{i.e.} the maximum difference between the up and down strain sweeps, shown here for $x=0.075$ at $T=70~$K in the transverse geometry. For a given $x$ value, the hysteresis width shown corresponds to the (transverse or longitudinal) measurement with the best signal to noise ratio. At any given temperature we report the value obtained from an average of several strain sweeps, the positive (resp. negative) error bars are extracted from the maximal (resp. minimal) value of individual strain sweeps. At temperatures where no hysteresis is resolved, the hysteresis width has to be smaller than the experimental noise and this is shown as a positive error bar extending down to zero. The arrows indicate the corresponding $T^*$ temperatures of maximum $m_{11}-m_{12}$ (see Fig.\ref{Fig3}). For $x=0.10$ no finite hysteresis is resolved down to the lowest temperature.}
    \label{Fig4}
\end{figure}

\subsection{Strain hysteresis} 
A peculiar feature of the reported elastoresistance is the presence of reproducible hysteresis in the resistance versus strain sweeps (see Fig.\ref{Fig1}c). This was previously reported in Ba$_{1-x}$Sr$_{x}$Ni$_2$As$_2$ and attributed to the pinning of static nematic domains by the I-CDW \citep{Eckberg_2020, Lee_PRL_2021}. Similar hysteresic behavior was also observed in $R$Te$_3$ ($R$= Tm, Er) and ascribed to a first-order reorientation of the CDW wavevector with uniaxial stress \citep{Straquadine_Arxiv_2020}.

In Fig.\ref{Fig4} we report the evolution of the hysteresis width, \textit{i.e.} the maximum difference of the relative resistance variation between the up and down strain sweeps (see inset), as a function of P-substitution and temperature. In all samples at high temperature no hysteresis is seen, \textit{i.e.} the response of electrical resistance to strain appears perfectly reversible. In that case, we can only define an upper limit on the unresolved and/or non-existent hysteresis width based on the experimental noise. This is depicted as vertical error bars extending to zero. Below a substitution-dependent temperature a finite hysteresis is resolved, meaning in particular that the hysteresis width is larger than the experimental noise. Note that only for $x=0.10$ no finite hysteresis is resolved down to the lowest temperature (see SM section IV).

For all other substitution levels we observe a rather sharp increase of the hysteresis width across the $T^*$ temperature of maximum elastoresistance (see arrows). While for BaNi$_2$As$_2$ a finite hysteresis is found to emerge at $T \approx T_\textrm{I-CDW}$ within resolution, in agreement with Ref. \citep{Eckberg_2020}, this is not the case in P-substituted samples. Since the hysteresis width significantly increases across $T^*$ for $0 \leq x \leq 0.075$ one likely scenario is that the hysteresis is associated to the pinning of orthorhombic domains. This is consistent with the orthorhombic domains orientation \citep{Merz_PRB_2021}. Testing this scenario would require measurements under larger strain and a precise knowledge of the temperature and substitution dependencies of the spontaneous orthorhombic distortion \citep{Hicks_PRX_2021}. The remarkable absence of a finite hysteresis at $x=0.10$ might be a signature of the absence of the orthorhombic distortion, in line with the absence of the associated detwinning effect seen in thermal expansion measurements \citep{Meingast_InPreparation}. Finally, the hysteretic strain behavior persists into the triclinic phase where the associated structural and/or C-CDW domains probably play a major role.

\begin{figure}
    \centering
    \includegraphics[width=8.6cm]{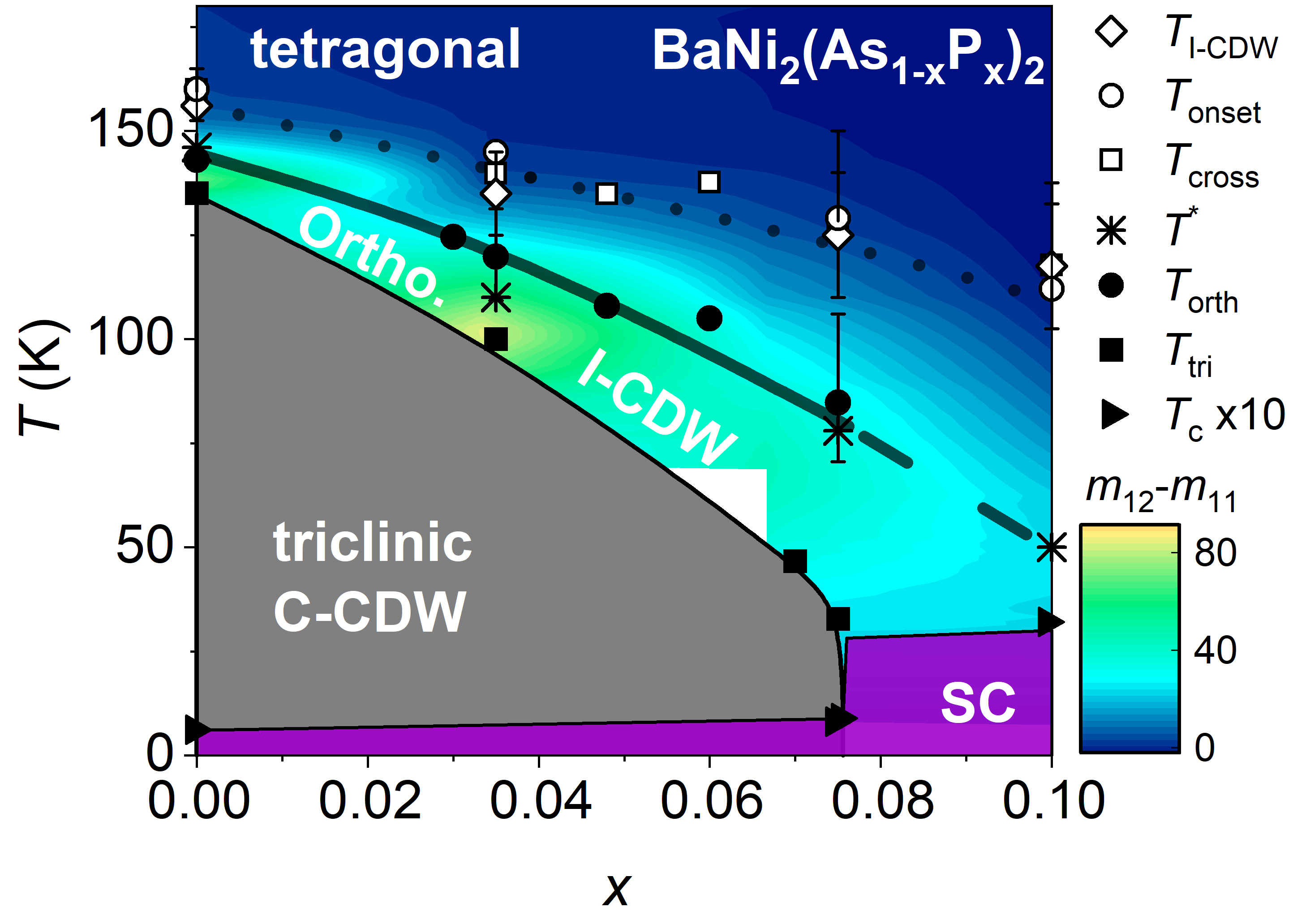}
    \caption{\textbf{Phase diagram of BaNi$_2$(As$_{1-x}$P$_x$)$_2$.} Colour map phase diagram showing the ($B_{1g}$) $m_{12}-m_{11}$ elastoresistance coefficient. The onset of the elastoresistance signal, $T_{\textrm{onset}}$ (empty circles), is in good agreeement with $T_\textrm{I-CDW}$ (empty diamonds), corresponding to a strong increase in the I-CDW superlattice peak intensity as determined by x-ray experiments \citep{Raman_KIT_Preprint}, and $T_{\textrm{cross}}$ (empty squares) that denotes a maximum in the $c/a$ ratio \citep{Meingast_InPreparation}. $T_{\textrm{c}}$ is the superconducting critical temperature as determined by specific-heat \citep{Raman_KIT_Preprint} (multiplied by a factor 10 for clarity; triangles, magenta area). $T^*$ (stars) denotes the $m_{12}-m_{11}$ maximum, in good agreement with $T_\textrm{orth}$ as determined by thermal expansion (closed circles, from Ref. \citep{Meingast_InPreparation}). $T_{\textrm{tri}}$ (closed squares) is the triclinic transition temperature determined from freestanding resistance measurements upon cooling. Lines are guide to the eye. } 
    \label{Fig5}
\end{figure}

\section{Discussion}

We summarize our results in a phase diagram, see Fig. \ref{Fig5}. A large $m_{12}-m_{11}$ elastoresistance coefficient onsets together with the I-CDW phase and has a maximum at the orthorhombic transition up to $x=0.075$. Remarkably, this elastoresistance onset also corresponds to a maximum in the crystallographic $c/a$ ratio which occurs at the $T_{\rm{cross}}$ temperature reproduced from Ref. \citep{Meingast_InPreparation}. At $x=0.10$, even though no orthorhombic transition has been reported by high-resolution thermal expansion \citep{Meingast_InPreparation} a weaker and broader $m_{12}-m_{11}$ maximum is found. In parallel, the hysteretic behavior of the elastoresistance reported at lower substitution levels is not resolved anymore. Notably, upon increasing substitution level, the enhanced superconducting $T_\textrm{c}$ above the triclinic critical point coincides with a reduction of $m_{12}-m_{11}$, and not an enhancement. Together with the already $T_{\rm{c}} \approx 3~$K superconductivity of fully substituted BaNi$_2$P$_2$ \citep{Mine_SolidStateComm_2008}, this observation strongly suggests that the superconductivity of BaNi$_2$(As$_{1-x}$P$_x$)$_2$ with $x>x_{\textrm{c}}$ is not significantly boosted by electronic nematicity.

We now consider in more details the elastoresistance signal itself. The advent of a large $B_{1g}$ elastoresistance response maximized across the tetragonal-to-orthorhombic transition of BaNi$_2$(As$_{1-x}$P$_x$)$_2$ is certainly reminiscent of the electronic nematic transition of its iron-counterpart, BaFe$_2$As$_2$. However, there are fundamental differences between these two cases, which are highlighted in the following.

In the iron pnictides the $m_{66}$ elastoresistance coefficient, probing nematicity in the $B_{2g}$ channel, follows a typical Curie-Weiss dependence over a wide temperature range of up to 100 K or more above the orthorhombic transition. Such a temperature dependence is expected on theoretical grounds for an electronically driven nematic transition \citep{Chu_Science_2010, Chu_Science_2012, Palmstrom_PRB_2017, Meingast_NatComm_2015}. This temperature dependence is also observed above the nematic transition of FeSe$_{1-x}$S$_x$  whose spin and/or orbital origin is still debated \citep{Meingast_PRB_2018, Hosoi_PNAS_2016, Watson_PRB_15}. In sharp contrast, in BaNi$_2$As$_2$, $m_{12}-m_{11}$, probing nematicity in the $B_{1g}$ channel, increases only in the close vicinity of the orthorhombic transition and its onset corresponds to a strong increase in the I-CDW superlattice peak intensity seen in x-ray diffraction experiments. The difference is evident in the comparison shown in the inset of Fig.\ref{Fig3}a where BaFe$_2$As$_2$ is choosen as it shares a close orthorhombic transition temperature. While for the latter compound the $m_{66}$ elastoresistance coefficient can be tracked up to $150~$K above the nematic transition, for BaNi$_2$As$_2$ $m_{12}-m_{11} \sim 1$ only $20~$K above $T_\textrm{orth}$. Hence, in BaNi$_2$As$_2$ the temperature dependence of the elastoresistance is very different from that of the well-established electronic nematic systems and in particular does not exclude a purely lattice-driven structural transition \citep{Kuo_Science_2016, Chu_Science_2012}.

With increasing substitution level, the $m_{12}-m_{11}$ maximum broadens, which can be ascribed to the broadening of the orthorhombic distortion seen in thermal expansion \citep{Meingast_InPreparation}, the increasing disorder, and finally the effect of external stress applied across the orthorhombic transition from the thermal expansion mismatch with the piezo. The latter effect is significant when the externally applied strain is of the same order as the spontaneous orthorhombic distortion. We expect this situation to be realized in BaNi$_2$(As$_{1-x}$P$_x$)$_2$ where, within our experimental conditions, both quantities are $\sim 10^{-4}$ \citep{Merz_PRB_2021, Raman_KIT_Preprint, Meingast_InPreparation} (see thermal expansion measurements in SM section II). In addition, as for $x=0$, $m_{12}-m_{11}$ is negligible above $T_\textrm{I-CDW}$, advocating for a similar origin. Finally, even in substituted samples, a Curie-Weiss susceptibility cannot fairly describe $m_{12}-m_{11}$ over a significant temperature range. Thus, in pure and P-substituted samples, the large $m_{12}-m_{11}$ elastoresistance coefficient is not an evidence for electronic nematicity, but is a property of the I-CDW phase. We note that the absence of a significant critical electronic nematicity is consistent with Young modulus measurements at $x=0.10$ \citep{Meingast_InPreparation} and an ARPES study of BaNi$_2$As$_{2}$ under uniaxial stress \citep{Guo_ArxiV_2022}.

 Let us now consider in more details the parallel strong increase of the I-CDW superlattice peak intensity and the onset of the $m_{12}-m_{11}$ elastoresistance coefficient. Regardless of the exact mechanism of formation of the I-CDW, its existence points to a significant coupling between the lattice and electronic degrees of freedom \citep{Johannes_PRB_2008}. Perturbation of the charge ordered state can then be achieved when strain of the right symmetry is applied \citep{Straquadine_Arxiv_2020} and a corresponding electrical transport signature is expected through, for instance, modification of the Fermi surface \citep{Sinchenko_PRL_2014}. Consequently, our observation of a large $m_{12}-m_{11}$ elastoresistance coefficient only for $T \leq T_\textrm{I-CDW}$ points towards a coupling between the I-CDW order parameter, the associated fluctuations, and the $\epsilon_{xx}-\epsilon_{yy}$ strain. In turn this observation strongly suggests an intimate relationship with the orthorhombic distortion, that manifests itself as a $m_{12}-m_{11}$ maximum, and that we interpret as a signature of the long-range order of the I-CDW. Importantly, this contrasts with Ba$_{1-x}$Sr$_{x}$Ni$_2$As$_2$ where, for $x \gtrsim 0.5$, a large $m_{12}-m_{11}$ elastoresistance coefficient occurs in the absence of the I-CDW phase and where, additionally, the fate of the orthorhombic phase remains to be investigated \citep{Eckberg_2020, Lee_PRL_2021, Lee_PRL_2019}. Our results, in particular the significant sensitivity of the I-CDW to uniaxial stress, put strong constrains on the theoretical description of this phase. It should motivate a refinement of its real space structure, and spectroscopic studies under uniaxial stress. Another promising avenue for a deeper understanding of the exact relationship between I-CDW, elastoresistance and structural distortion is the recently developed \textit{in-situ} combination of strain-dependent x-ray diffraction and electrical transport measurements \citep{Sanchez_NatMat_2021}.

Finally, we focus on a particularly intriguing case, BaNi$_2$(As$_{0.9}$P$_{0.1}$)$_2$, located above the triclinic critical point, $x_c$. As for lower substitution levels, the onset of $m_{12}-m_{11}$ does correspond to a strong increase in the I-CDW superlattice peak intensity. However, as shown in Fig.\ref{Fig4} a strain hysteresis is not resolved at this substitution content anymore. In parallel, no detwinning of the orthorhombic domains is seen in high-resolution thermal expansion measurements, questioning the occurrence of an orthorhombic phase transition for this composition. A first-order transition is observed in thermal expansion within the temperature range of the $m_{12}-m_{11}$ maximum \citep{Meingast_InPreparation}, but the absence of a thermal hysteresis in the elastoresistance points toward a different origin. Moreover, the resistance measurements do not reveal any evidence of a phase transition in the vicinity of $T^* \approx 50~$K (see SM section V). Noteworthy, the onset of the $m_{12}-m_{11}$ elastoresistance coefficient coincides with a lattice softening as seen by Young modulus measurements, which is argued to be incompatible with critical electronic nematicity \citep{Meingast_InPreparation}. The simultaneous maximum of elastoresistance and saturation of the softening below $\approx 50~\textrm{K}$ strongly suggests a close relationship between those two. The $m_{12}-m_{11}$ maximum, though broad, in the (likely) absence of an orthorhombic transition calls for further investigations.

In conclusion, we report a large $B_{1g}$-symmetric $m_{12}-m_{11}$ elastoresistance coefficient in the close vicinity of the tetragonal-to-orthorhombic transition of BaNi$_2$(As$_{1-x}$P$_x$)$_2$. While the observation of an elastoresistance maximum at this structural transition is certainly reminiscent of the iron pnictides, the temperature dependence of $m_{12}-m_{11}$ strikingly contrasts with known examples of electronic nematic transitions. In particular, it does not follow the typical Curie-Weiss-like form. Rather, the strong increase in the I-CDW satellite intensity observed in parallel to the onset of the elastoresistance signal indicates that the strain-dependent electrical transport is a property of the I-CDW phase. The weakening of the $m_{12}-m_{11}$ elastoresistance coefficient observed in parallel to the enhanced superconducting $T_{\textrm{c}}$ in the absence of the triclinic structure strongly suggests that the strain-sensitive electronic correlations revealed by elastoresistance are not responsible for the stronger superconductivity. Finally, a careful inspection of the hysteretic behavior of the resistance versus strain sweeps points to the pinning of orthorhombic domains as a likely origin.

\section{Methods}

\subsection{Single crystals growth and chemical analysis}
Single crystals of BaNi$_2$(As$_{1-x}$P$_x$)$_2$ (with $x= 0,~ 0.035,~ 0.07,~0.075, ~0.10$) were grown using a self-flux method. NiAs binary was synthesised by mixing the pure elements Ni (powder, Alfa Aesar 99.999$\%$) and As (lumps, Alfa Aesar 99.9999$\%$) that were ground and sealed in a fused silica tube and annealed for 20 hours at 730 °C. All sample handlings were performed in an argon glove box (O$_2$ content $<$ 0.5 ppm). For the growth of BaNi$_2$(As$_{1-x}$P$_x$)$_2$, a ratio of Ba:NiAs:Ni:P = 1:4$(1-x)$:4$x$:4$x$ was placed in an alumina tube, which was sealed in an evacuated quartz ampule (i.e. $10^{-5}$ mbar). The mixtures were heated to 500°C-700°C for 10 h, followed by heating slowly to a temperature of 1100°C-1180°C, soaked for 5 h, and subsequently cooled to 1000°C-900°C at the rate of 0.5°C/h to 2 °C/h, depending on the phosphorus content used for the growth. At 1000°C-900°C, the furnace was canted to remove the excess flux, followed by furnace cooling. Plate-like single crystals with typical sizes 3 x 2 x 0.5 mm$^3$ were easily removed from the remaining ingot. The crystals were brittle having shiny brass-yellow metallic lustre. Electron micro probe analysis of the BaNi$_2$(As$_{1-x}$P$_x$)$_2$ crystals was performed using a compact scanning electron microscope (SEM) – energy dispersive x-ray spectroscopy (EDS) device COXEM EM-30plus equipped with an Oxford Silicon-Drift-Detector (SDD) and AZtecLiveLite-software package. The EDS analyses on the BaNi$_2$(As$_{1-x}$P$_x$)$_2$ crystals revealed phosphorus content  $x= 0,~ 0.035,~ 0.070,~0.075$ and $0.10$ with a typical uncertainty of $\Delta x = \pm 0.05$.

\subsection{Elastoresistance measurements}
DC-elastoresistance measurements were performed following the method described in Ref. \citep{Kuo_PRB_2013}. We used piezoelectrics from Piezomechanik GmbH (Part. No. Pst 150/5x5x7) and miniature strain gauges from Vishay Precision Group. The samples and strain gauges were glued to opposites faces of the piezo using DevCon 5mn 2-components epoxy (Part. No. X0039). Bias voltages from +150V to -30V were used. The maximal strain $\epsilon_{xx}$ applied along the piezo poling direction decreases with decreasing temperature, going from $\sim 0.1 \%$ at 300~K to less than $\sim 0.03 \%$ at $50~$K. The experimental setup has been checked and validated by measuring the well-known longitudinal response of BaFe$_2$As$_2$ \citep{Chu_Science_2012}. To extract the symmetry-resolved $m_{11}-m_{12}$ elastoresistance coefficient we consider a temperature independent piezo stack Poisson ratio $\nu = - \epsilon_{yy}/\epsilon_{xx} \approx 0.43$ \citep{Kuo_PRB_2013}.

The single crystals were cut with edges along the [100]$_\textrm{tet}$ direction, with typical dimensions 1.5mm $\times$ 1mm $\times$ 50 $\mu m$. The small thickness along the [001]$_\textrm{tet}$ direction is necessary for high and homogeneous strain transmission. The strain value extracted from the strain gauge is considered as the strain felt by the sample. A standard 4-contact geometry was used to measure the electrical resistance along the [100]$_\textrm{tet}$ direction using a Lake Shore 372 resistance bridge. Some of the freestanding resistance measurements were also done using a combination of Keithley 6221 current source and Keithley 2182A nanovoltmeter in delta mode. Either DuPont 4929N or Hans Wolbring Leitsilber silver paints were used.

Several strain sweeps were performed at each temperature to ensure the reproducibility of the extracted elastoresistance coefficients. A slow rate of 6V/s was used to drive the piezo. If possible cooling and warming elastoresistance measurements were recorded, without any significant difference. All the presented elastoresistance coefficients are extracted from linear-in-strain fits. Second-order fits do not lead to any significant change in the extracted first-order coefficients.

\section{References}

\bibliography{Elasto_BNAP}

\section{Acknowledgments}
We thank R. Willa, I. Vinograd and F. Hardy for valuable discussions. We acknowledge support by the Deutsche Forschungsgemeinschaft (DFG; German Research Foundation) under CRC/TRR 288 (Projects B03 and A02) and the Helmholtz Association under Contract No. VH-NG-1242. \mbox{M. F.} acknowledges funding from the Alexander von Humboldt foundation and the Young Investigator Group preparatory program of the Karlsruhe Institute for Technology. K. W. acknowledges funding from the Swiss National Science foundation through the postdoc mobility fellowship. S. M. S. acknowledges funding by the Deutsche Forschungsgemeinschaft (DFG, German Research Foundation) – Projektnummer 441231589.\\


\section{Competing interests}
The authors declare no competing interests.

\end{document}